\documentclass[%
 aip,
rsi,%
 amsmath,amssymb,
 reprint,%
]{revtex4-2}

\usepackage{graphicx}% Include figure files
\usepackage{dcolumn}% Align table columns on decimal point
\usepackage{bm}% bold math

\usepackage{listings}

\usepackage{verbatim}
\usepackage{fancyvrb}
\usepackage[dvipsnames]{xcolor}

\usepackage{hyperref}
\usepackage{cleveref}

\newcommand{\derp}[2]{\frac{\partial#1}{\partial#2}} \newcommand{\dderp}[2]{\frac{\partial^2#1}{\partial#2^2}} \makeatletter \newcommand{\pushright}[1]{\ifmeasuring@#1\else\omit\hfill$\displaystyle#1$\fi\ignorespaces} \newcommand{\pushleft}[1]{\ifmeasuring@#1\else\omit$\displaystyle#1$\hfill\fi\ignorespaces} \makeatother

\begin{document}

\preprint{AIP/123-QED}

\title[Rigo, Biau \& Gloerfelt]%{ A simplified approach to the curved duct flow, \\from stationarity to chaos}
{ Bifurcations in curved duct flow \\based on a simplified Dean model}
\author{Leonardo Rigo}
\author{Damien Biau}%
 \email{damien.biau@ensam.eu}

\author{Xavier Gloerfelt}
 %\homepage{http://www.Second.institution.edu/~Charlie.Author.}
\affiliation{ 
Dynfluid Laboratory, \'{E}cole Nationale Superieure d'Arts et M\'etiers.\\
151 Boulevard de l'H\^opital, 75013 Paris, France%\\This line break forced with \textbackslash\textbackslash
}%
%\affiliation{%
%Second institution and/or address%\\This line break forced% with \\
%}%

\date{\today}% It is always \today, today,

\begin{abstract}
We present a minimal model of an incompressible flow in  square duct subject to a slight curvature. 
Using a Poincar\'e-like section we identify stationary, periodic, aperiodic and chaotic regimes, depending on the unique  control parameter of the problem: the Dean number ($De$).
Aside from representing a simple, yet rich, dynamical system the present   simplified model is also representative of the full problem, reproducing quite accurately the bifurcation points observed in the literature. 
We analyse the bifurcation diagram from $De=0$ (no curvature) to $De=500$,  observing a periodic segment  followed by two separate chaotic regions. 
The phase diagram of the flow in the periodic regime shows the presence of two symmetric steady states, the system oscillates around these  solutions following a  heteroclinic cycle.
In the appendix some quantitative results are provided for validation purposes, as well as the python code used for the numerical solution of the Navier-Stokes equations.

%Valid PACS numbers may be entered using the \verb+\pacs{#1}+ command.
\end{abstract}

\pacs{Valid PACS appear here}% PACS, the Physics and Astronomy
\keywords{Navier-Stokes equations, deterministic chaos, curved flow, Dean vortices  }%Use showkeys class option if keyword
\maketitle

\begin{quotation}
The flow in a curved square duct is governed by the Navier-Stokes (NS)  equations  in polar coordinates. The dynamics of this kind of flow can be rather complex, with stationary, periodic or chaotic regimes depending on the flow parameter, namely the Reynolds number, and the non dimensional curvature of the duct. Starting from these equations we derive a minimal model for this problem, in line with that introduced by \citet{dean1927,Dean1928,dean_hurst_1959}. The numerical  solution of the new set of equations is drastically simplified with respect to the polar NS equations, but the richness of the problem  is  kept. The present model, aside from being a simplified tool to calculate the flow in a bent duct, is an interesting dynamical system ranging from stationarity to periodicity, aperiodicity and chaos, all depending on a single control parameter, namely the Dean number.

% The ``lead paragraph'' is encapsulated with the \LaTeX\ 
% \verb+quotation+ environment and is formatted as a single paragraph before the first section heading. 
% (The \verb+quotation+ environment reverts to its usual meaning after the first sectioning command.) 
% Note that numbered references are allowed in the lead paragraph.
% %
% The lead paragraph will only be found in an article being prepared for the journal   \textit{Chaos}.
\end{quotation}

\section{Introduction}

%Internal flows in straight or bent ducts are found in countless industrial applications, from HVAC systems to cooling applications in turbomachinery or nuclear plants, in this paper we will focus on ducts of unitary aspect ratio (square cross section). The flow in a straight square duct bears significant analogies with pipe flow, the two laminar velocity profiles are quite similar and both flows are linearly stable for any value of the Reynolds number $Re$, as shown by \cite{tatsumi_yoshimura_1990} and \cite{drazin1981hydrodynamic}). The introduction of a curvature significantly changes this behaviour: the critical $Re$, which in a straight pipe is around $2040$ according to \cite{AvilaScience2011}, can be increased in a curved pipe, to prove this \cite{viswanath2013visualization} showed that in a helically coiled pipe followed and preceded by a straight segment, a turbulent flow entering the first straight section can be observed to relaminarize in the coiled part, and then transition again to turbulence  in the following straight parcel. This is in agreement with \cite{goertler1940} who observed a similar phenomenon in a curved plate case. 

\label{sec:introduction}
%The flow in a curved duct of square or round cross section present
The appearance of two symmetric counter-rotating secondary vortices is a well known phenomenon in the laminar curved pipe and square duct flows, which show qualitatively identical behaviours.  The first experimental observations date back to the beginning of the $20^{th}$ century   \citep{Eustice1910}. A more quantitative understanding of the problem was provided soon after by 
\citet{dean1927,Dean1928} and later by \citet{dean_hurst_1959}, who proposed  a simplified model for the flow in a pipe or square duct subject to a slight curvature, based on a truncation of the Navier-Stokes (NS) equations written in polar coordinates.  The model depends on a single control parameter (that has been known ever since as the Dean number -$De$-), it is two-dimensional and stationary. It provided a first insight on the structure of the counter rotating vortices and their effects  on this type of flows, especially on the pressure drop. Despite some further  analytical studies \cite{McConalogueSrivastava1968,Vandyke_1978} most of the works that followed  \cite{cheng1969laminar,Joseph1975,Cheng1976,dennis1982dual,daskopoulos1989flow,YANASE_dualsolutionsPipe_1989}  relied no longer on Dean's approximations.  The rapidly developing computational resources allowed for the solution of the complete NS equations in two dimensions (2D), also taking  advantage of the  geometrical lateral symmetry of the problem. These studies showed the presence of an  additional couple of counter-rotating vortices, produced by the G\"ortler instability on the concave wall, situated at the outer part of the duct. This instability appears for high enough curvature and flow velocity through a subcritical  bifurcation \cite{Shanthini1986}, this subcriticality is strongly linked to the arbitrary imposed  lateral symmetry.
In particular \citet{Cheng1976} performed a parametric study for different Dean numbers showing that for increasing $De$ the flow evolved from 2 to 4 vortices, sometimes referred to as circulation cells, returning to a 2 cells structure and subsequently back  to  4 cells, hence bearing a non trivial dependence on $De$. For a more extensive review on the early works on the subject see Ref \onlinecite{ito1969review, berger1983Review}. 
%Extensive experimental works were also conducted, confirming the presence of these solutions .

\citet{winters1987bifurcation} largely contributed to the understanding of this problem, specifically referring to the square duct case, clarifying some important aspects. He gave a quantitative criterion for the distinction between low and high curvature, showing that for $h/R>0.05$ ($h$ and $R$ being the duct height and the curvature radius respectively) 
the bifurcation scenario is governed by two independent parameters, the Reynolds number $Re$ and the geometry $h/R$, whereas for $h/R<0.05$ the curvature can be considered as weak and the dynamic is governed by a single control parameter, $De=Re\sqrt{h/R}$. In addition, he shed light on the effect of the symmetry hypothesis, which was found to artificially stabilise the 4-vortices solution, thus altering the 2 to 4-cells bifurcation, which was earlier deemed subcritical \cite{Shanthini1986} but is in fact a supercritical pitchfork bifurcation.  Most importantly \citet{winters1987bifurcation} performed a parametric study by varying the two control parameters. He presented,  for the first time, a bifurcation diagram above the steady state originally obtained by \citet{dean1927} 60 years before. 
An open question remains: is it possible to retrieve the results of \citet{winters1987bifurcation} using the   Dean simplification? Indeed, a valid simplified model would have a great interest for more complex flows.
%,  \citet{Wang2004} 
%After the first bifurcation ($De=128.22$)\cite{Wang2004} no stable branches are present, the system periodically  oscillates between two unstable fixed points, re-becoming stationary between  $De=217$ and $310$ where a linearly stable branch is again  present\cite{Wang2004,Wang2005}. For increasing $De$ the system becomes aperiodic and eventually chaotic \cite{Wang2004,mondal2007}. 
%\citet{Wang2005} DICONO CHE 2D OK ANCHE PER PERIODICOshowed that the 2D model correctly represents the fully developed experimental flow in the stationary regime, after the bifurcation the behaviour is better described by a   

As an example of complexity, turbulent flows in duct or pipe are impacted by curvature.
%Curvature  has a relevant impact on turbulence in pipe and square duct flows.
 \citet{white_1929, taylor1929}  experimentally demonstrated that it has a stabilising effect, increasing the critical Reynolds number with respect to the straight case. \citet{Sreenivasan1983} showed that a turbulent flow entering in a coiled section becomes laminar and becomes again turbulent in the subsequent straight section.  Recently the laminar to turbulent bifurcation was shown to change from subcritical, in the straight case, to supercritical for large enough curvatures, with a   smooth transition to turbulence \cite{rinaldi_canton_schlatter_2019,canton_schlatter_orlu_2016,kuhnen_braunshier_schwegel_kuhlmann_hof_2015,kuhnen_holzner_hof_kuhlmann_2014,barkley_2019}. For a complete review on the effects of curvature on turbulence the reader is referred to the recent review by \citet{vester_2016Review}.

Whether or not a  2D model can be representative of the real flow even in the non stationary regimes is debatable. Some authors\cite{mondal2007,Wang2005,YanaseWatanabe2008} assert that this is the case at least in the periodic regime, implying the presence of travelling waves, which were indeed  experimentally observed by \citet{Mees1996_steadyOscill,mees_nandakumar_masliyah_1996_sec_instab}. In any case the 2D curved duct problem is per-se an intriguing dynamical system, with several bifurcations depending, formally,  on  two control parameters, characterising the flow and the geometry. 

In this paper we propose a more extensive use of  Dean's  model, which is modified correcting an inconsistency present in  original formulation by \citet{dean1927,Dean1928,dean_hurst_1959}.
Moreover it is time dependent and 2D in this work, but can be extended to 3D cases.
%Moreover it is time dependent and three dimensional, although it will be used in 2D in this work.
Alike in Dean's model there is one single control parameter ($De$) and the problem is in  Cartesian coordinates, which simplifies the numerical solution. Despite the simplicity of this formulation the complexity of the problem is preserved, the bifurcations are recovered depending only on $De$. 
%with stationary, periodic, aperiodic and eventually chaotic behaviours, this time  depending only on $De$.  
The outline of the paper is as follows: in section \ref{sec:governing_eqns} we introduce the simplified model, clarifying the hypotheses that allow its derivation from the NS equations in polar coordinates and the scaling law between the reference quantities. Section \ref{sec:bifurc_diagram} will give a broad view of the bifurcation scenario that characterises the problem, we will introduce the methodology used to calculate the time periods and we will compare the critical Dean numbers given by the model with  literature results. We will then focus on the periodic regime, analysing the dependence of the oscillation period on the Dean number, and then analysing the spectrum of time series and the periodic orbit at  $De=130$. We will then analyse the behaviour at high Dean number -$De=10^4$- which shows spatio- temporal complexity, with vortices present in the whole domain, contaminating the 4 walls, with a multi-scale distribution. In appendix  we  provide  details about the program used in this article,  the script is made  available in the last section.

\section{Governing equations\label{sec:governing_eqns}}
We start from the incompressible Navier-Stokes equations in polar coordinates. The  nabla operator in 2D polar coordinates  is 
\begin{equation}
\label{eqn:nabla_polar}
\nabla:=
   \left(\begin{array}{cc}
     %\frac{1}{R+r} \derp{}{\theta} \\ 
     %\\
     \derp{}{r}+\frac{1}{R+r} \\
     \\
     \derp{}{z}
   \end{array}
   \right)
\end{equation}
where $r,z$ are the radial and axial coordinates respectively and $R$ is the centerline curvature radius.
In the following $(x,u)$; $(y,v)$; $(z,w)$  denote the local Cartesian system in streamwise, radial and cross stream directions and velocities, respectively, with
\begin{eqnarray*}
y:=r+R \approx R \\ \nonumber 
dy=dr \nonumber \nonumber
%x:=\left(r+R \right)sin(\theta)\approx R\theta
\end{eqnarray*}
%$y$ being the local radial coordinate.   

If we assume that the non dimensional  curvature is  weak , namely $h/R\ll1$ with $h$ being the duct height,  then the global polar coordinates can be approximated by the local cartesian coordinates. 

The simplified model by  \citet{dean1927,Dean1928,dean_hurst_1959} is based on the hypothesis that  $h/R\ll 1$ as in the present paper, but using one characteristic velocity. By truncating the equations at order $h/R$ Dean neglected most of the additional terms  in the  equations in polar coordinates, but  arbitrarily kept the Coriolis acceleration, which is formally also of order $h/R$.
Thus this term should be neglected, but in that case we would suppress any curvature effect.
This issue can be overcome by using two different velocity scales 
%In order to build a minimal model for curvature it is necessary to use 
%We start from the incompressible Navier-Stokes equations in polar coordinates. $(x,u)$; $(y,v)$; $(z,w)$  denote the streamwise, radial and cross stream directions and velocities.
%The non dimensionalisation is performed using 
%two reference velocities 
$U_b,V_0$ for streamwise  and spanwise  directions, $U_b$ is the streamwise bulk velocity
\begin{equation}
    U_b:=\int_S U~ dS.
\end{equation}
By doing so, we build a minimal model, the curvature effect being reduced to a single term in a cartesian framework: the Coriolis acceleration $-u^2/R$.
The relationship between the reference quantities stems from the assumption
that the order of magnitude of the centrifugal term $u^{2}/R$ in the momentum equation for $y$ is  the same as that of the non linear terms. We have 
\begin{equation}
    \label{eqn:scaling_law}
    \frac{V_0^2}{h}=\frac{U_b^2}{R}  
    %\hspace{1cm} 
    \implies
    \frac{V_0}{U_b}=\sqrt{\frac{h}{R}}\hspace{1cm}
    %\frac{U_bV_0}{L}=\frac{U_b^2}{R}
\end{equation}
By injecting this scaling law in the equations and neglecting the terms of  $\mathcal{O}(h/R)$ or higher, we obtain 
\begin{subequations} 
\label{eqn:NS_NONdimensional_minimalmodel_xyz}
\begin{alignat}{5}
  \label{eqn:continuity_NONdimensional_minimalmodel_xyz}
    & \derp{v}{y}   +\derp{w}{z}=0   \\
    \label{eqn:momentum_u_NONdimensional_minimalmodel_xyz}
    &\derp{u }{t }+De\left(  v \derp{u }{y }+w \derp{u }{z}      \right)&&=-\frac{dP}{dx}  +     &&\dderp{u}{y}+ \dderp{u}{z}   \\ \label{eqn:momentum_v_NONdimensional_minimalmodel_xyz}
    &\derp{v }{t }+De\left(  v \derp{v }{y }+w \derp{v }{z} -\mathbf{u^2} \right)&&=-\derp{p}{y} +   && \dderp{v}{y}+ \dderp{v}{z}  \\ 
    \label{eqn:momentum_w_NONdimensional_minimalmodel_xyz}
    &\derp{w }{t }+De\left(  v \derp{w }{y }+w \derp{w }{z}      \right)&&=-\derp{p}{z}+&&\dderp{w}{y}+ \dderp{w}{z}   \\
     \nonumber
\end{alignat}
\end{subequations}
Note that the additional term for the curvature ($-De~ u^2$) has been highlighted with a bold character.  %Where 
\begin{equation}
    De=Re\sqrt{\frac{h}{R}},\hspace{0.5cm} \mathrm{with} \hspace{0.5cm} Re=\frac{U_b h}{\nu}
    \label{eqn:Dean_Reynolds_definition}
\end{equation}
are  respectively  the Dean  and Reynolds numbers and
\begin{align}
u=\frac{u^*}{U_b}&\hspace{1cm}v,w=\frac{v^*,z^*}{V_0}\hspace{1cm} 
p=\frac{p^*h}{\mu V_0} \nonumber \\ 
\label{eqn:nonDim_qnt}
&t=\frac{t^*U_b^2}{\nu}\hspace{1cm}y,z=\frac{y^*,z^*}{h}\hspace{1cm}%x=\frac{x^*}{L}
\end{align}
are the non-dimensional variables, with $^*$ denoting the dimensional quantities. 

The pressure gradient $dP/dx$ in the streamwise momentum equation \ref{eqn:momentum_u_NONdimensional_minimalmodel_xyz} is the forcing term  allowing the fluid motion. It must be distinguished from the   hydrodynamic  pressure $p(y,z,t)$ present in the other equations.  

On the wall we impose the no-slip boundary conditions.
Details about the numerical method are provided in appendix \ref{sec:appendix} in addition with the python program.

The solutions can be distinguished according to their symmetry in the spanwise direction ($z$):
\begin{itemize}
\item symmetric solutions
\begin{eqnarray}
u(0.5+z)=u(0.5-z) \nonumber\\  v(0.5+z)=v(0.5-z) \nonumber\\ w(0.5+z)=-w(0.5-z) \nonumber
\end{eqnarray}
\item antisymmetric solutions
\begin{eqnarray}
u(0.5+z)=-u(0.5-z) \nonumber\\  v(0.5+z)=-v(0.5-z) \nonumber\\ w(0.5+z)=w(0.5-z) \nonumber
\end{eqnarray}
\end{itemize} 

% \subsection{The inconsistency in Dean's formulation}
% The simplified model by  \citet{dean_hurst_1959} is based on the hypothesis that  $h/R\ll 1$ like in the present paper, but using one characteristic velocity. By truncating the equations at order $h/R$ \citet{dean1927} neglect most of the additional terms  in the  equations in polar coordinates, but they arbitrarily keep the Coriolis acceleration, which is formally  of order $h/R$.
% This term should be neglected, but in that case we suppress any curvature effect.
% This inconsistency can be overcome by using two different velocity scales.
% %in the momentum equation in the radial direction $y$ (for the sake of clarity we use a notation  coherent with the rest of the present paper, which is not the same as that of the original work). 
% However, by performing a dimensional analysis we see that if we use only one characteristic velocity and one characteristic length the additional term in the continuity equation is of order $h/R$, like the  term $u^2/R$. Therefore one cannot  neglect $v/R$ in the continuity equation without also neglecting $u^2/R$ in the $y$ momentum equation, hence completely eliminating the curvature terms. Instead, by using two reference lengths and velocities (only one reference length in the 2D case) we solve this inconsistency: the $v/R$ term is $\mathcal{O}(h/R)$, whilst the $u^2/R$ term is $\mathcal{O}(\sqrt{h/R})$ (using the scaling law  (\ref{eqn:scaling_law})), and can be kept while the other is neglected.
% %\subsection{The numerical method}

\section{Results}
\label{sec:bifurc_diagram}
\subsection{Bifurcation diagram}
In order to determine whether the system is stationary, periodic, aperiodic or chaotic we use a Poincar\'{e} like section based on the three following criteria:
\begin{itemize}
    \item $v=0$
    \item $ \derp{v}{t}<0 $
    \item $w<0$
\end{itemize}
measured in a specific point  located at the centerline close to the outer wall, specifically at $y=0.9045,z=0.5$. % This point will be hereafter referred to as $p$.  
These conditions are  met only once per period, thus if the oscillation is periodic the system will cross the section every $T$ time units, and the numerical values will be the same. 
%moreover the crossing will happen in the same point of the section, hence  at these times the variables will assume the same values.

\begin{figure*}
\includegraphics[width=1\textwidth,keepaspectratio]{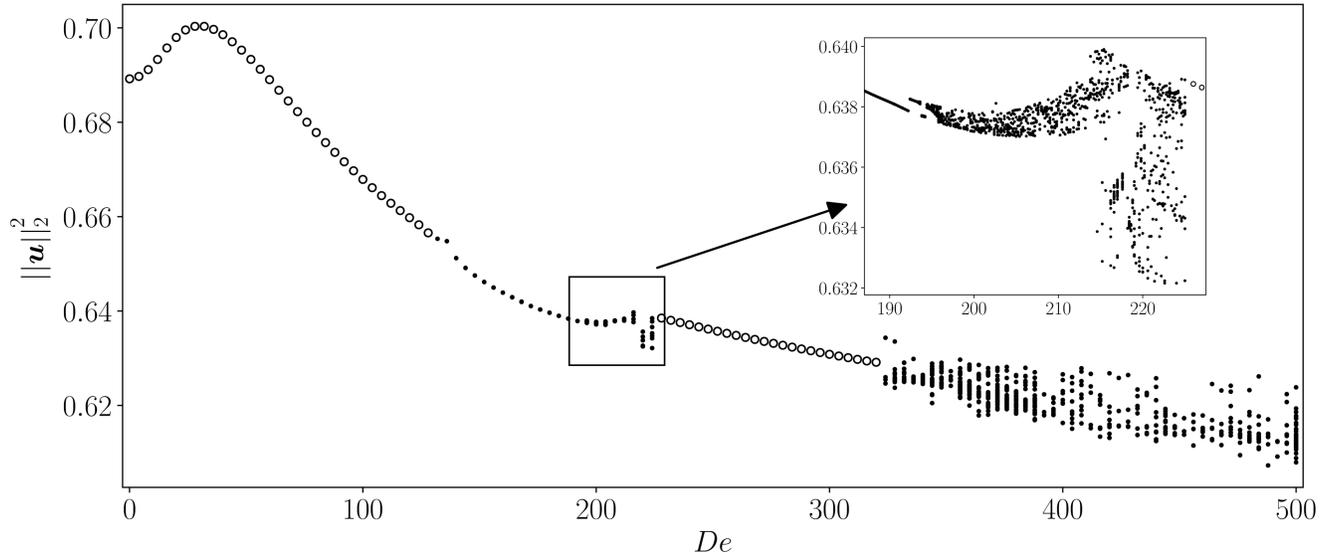}
\caption{\label{fig:bifurc_diagram} 
 Filled circles: values of the squared 2-norm of the velocity vector at the instants when the Poincar\'{e} section is crossed.  Open circles: values of the square of the 2-norm of the velocity vector in the stationary cases. 
 The first bifurcation from  a steady regime to an unsteady, periodic regime occurs at $De=128.32$. Starting from $De \approx 193$ we observe two periods which are incommensurable, characteristic of the aperiodic regime. Then the system  quickly evolves into chaos from $De \approx 195$. For  $De\in[227,318]$ the system is again stationary, after a    narrow periodicity interval the system becomes again chaotic at $De=324$. A magnification of the bifurcation diagram for $De\in[190,230]$ is shown 
 in the inset}
 %in the encapsulated figure }
\end{figure*}

% Figure \ref{fig:bifurc_diagram} was obtained  following a procedure similar to that used for figure \ref{fig:periodsPlot}, but instead of the oscillation period 
The open circles in figure \ref{fig:bifurc_diagram}  indicate the   values of the squared velocity 2-norm in the stationary regime. 
 The first bifurcation from  a steady regime to an unsteady, periodic regime occurs at $De=128.32$. This specific Dean number will be henceforth be referred to as the critical Dean number $De_c$. In the same figure, the black  dots represent the values of the squared 2-norm of the velocity vector at each crossing of the Poincar\'e-like section,  for each $De$ the simulation was run  for at least 10 periods. Hence the periodic behaviour is represented by at least 10 superposed points, as can be seen for instance at $De=130$.
 % bifurcation diagram FIGURE

%dot representing the velocity norm for instance at $De=130$ (where we have periodic oscillations, as we will see below) is not a single dot but instead the superposition of several  perfectly overlapping values. 
% The system is  stationary up to $De=128.32$, where a transition to a periodic regime occurs. 
%The low-$De$ region (for $De<De_c$) bifurcates into a periodic regime, 
By increasing the Dean number, we see that the system holds its periodicity until $De\approx193$ (using the definition of $De$ in equation \ref{eqn:Dean_Reynolds_definition}, which is not the same used in \citet{Wang2004,Wang2005}), where it becomes aperiodic. \citet{Wang2004,Wang2005}  also saw a change in behaviour in that Dean number range, a slight increase and then a decrease in the period, but they did not mention to observe aperiodicity.  %, however they did not push their analysis to higher Dean numbers. 
%seen in the last part of figure \ref{fig:periodsPlot}. 
This aperiodic regime is only observed over a narrow range of $De$,  already at $De=195$ the system is complex enough to be considered  as chaotic. 
Analysing the magnified part in figure \ref{fig:bifurc_diagram}, we notice the presence of a new solution branch to which  the system abruptly jumps back and forth. The increase in complexity is then very rapid, between $De=125$ and $214$ the system is chaotic with  values of the velocity norm that remain enclosed in the envelope of the two former branches. Between $De=214$ and $227$ the values  no longer lie within the same envelope, probably due to the presence of a third branch (which was indeed observed by \citet{Wang2004}). For $De>227$ the system very quickly returns to a stationary regime,  which lasts until $De=318$, where very mild, periodic oscillations are observed up to $De=323$. For higher Dean numbers the system rapidly becomes again chaotic, no stationary or periodic regimes were observed for higher Dean numbers.

%\subsection {Comparison of bifurcation points with literature results}
%{ Bifurcation points comparison with literature results  }
The changes in regime observed for the present model are in very good agreement with the literature. In particular the critical Dean number given by the model, $De_c=128.32$, is in excellent accordance with the $De_c=128.22$  by \citet{Wang2004}, which is one of the most accurate studies available, or  $De_c=129.71$ by \citet{winters1987bifurcation}. Both these results were obtained by solving the full  2D Navier Stokes equations in polar coordinates without low curvature approximation, thus with two control parameters. %\citet{sankar1988} obtained a value of $128.125$ using a 3D stationary, parabolic  formulation. 
The bifurcation to the new steady state occurs at $De=227$ with our model, which is slightly above the value obtained in Ref \onlinecite{Wang2004}, $De=217$.
The agreement is better for the next bifurcation to chaos, which  is observed at $De=323$ for both models. 

%The beginnings of the second stationary region at $De=217$ and of the chaotic section from $De=323$ were also observed in Ref \onlinecite{Wang2004}.
Nonetheless a difference is observed at the end of the stationary region: in the section between $De=318$ and $323$ our model produces slight periodic oscillations, while
%($v$ oscillates without becoming negative, $w$ is constant and equal to zero).
\citet{Wang2004}  report  an aperiodic oscillation from  $De=312$ until the beginning of the chaotic region at $De=323$.

\subsection{The periodic regime}
\begin{figure}
\includegraphics[width=0.45\textwidth,keepaspectratio]{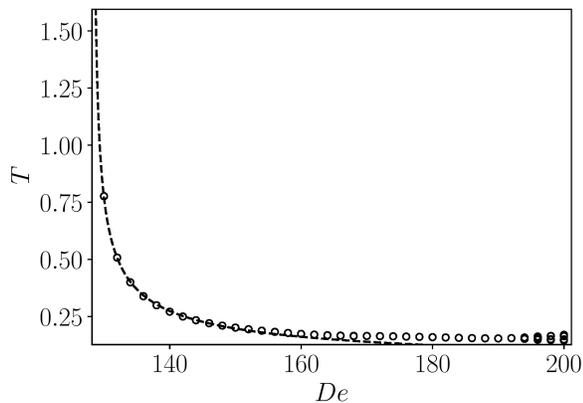}
\caption{\label{fig:periodsPlot} 
 The oscillation period follows the law: \newline $T\propto (De-De_c)^\gamma$,  represented by dashed line, with $\gamma=0.53$, and $De_c=128.32$. %The open circles indicate some of the calculated points. 
 Above $De\approx 193$ we observe the first part of aperiodic regime, which gives rise to higher complexity for higher $De$, see also figure \ref{fig:bifurc_diagram}.
 % dashed curve: 0.962/(De-128.371)**0.5
 }
\end{figure}

Figure  \ref{fig:periodsPlot} shows the dependence of the oscillation period on $De$ in the periodic regime. The open circles represent the time intervals between two crossings of the Poincar\'{e} section, i.e. the period. As for figure \ref{fig:bifurc_diagram}, the symbols are the superposition of several circles, onto one position for a periodic signal, and two or more in the aperiodic or chaotic regime.
At $De>193$ there is a progressive separation in two different set of circles, this identifies the aperiodicity seen in figure \ref{fig:bifurc_diagram}. 
\citet{Wang2004} report $T=0.159$ at $De=182.2$, retrieved with the full NS equations while  the present model gives $T=0.1579$ at the same $De$.
%The bifurcation from stationary to periodic regime observed at $De=128.32$ (this  Dean number will henceforth be referred to as the critical Dean number $De_c$ ) is in very good agreement with the value of $128.22$ reported by \citet{Wang2004} for the same bifurcation,  obtained by solving the full  2D Navier Stokes equations in polar coordinates. 
Between $De_c$ and $De=150$ the period dependence on $De$ follows the law:
\begin{equation*}
    T=\frac{1.013}{(De-De_c)^{0.53}}
\end{equation*} 
which
is represented with a dashed curve in the figure and 
is close to a $-0.5$ power law in agreement with supercritical pitchfork  bifurcation. In fact, the computed exponent approaches the theoretical value of $-0.5$ when the curve fitting is restricted to Dean numbers close to  the critical value, in agreement with the weakly non linear framework for the theoretical $-0.5$ value.

\begin{figure}
\includegraphics[width=0.45\textwidth,keepaspectratio]{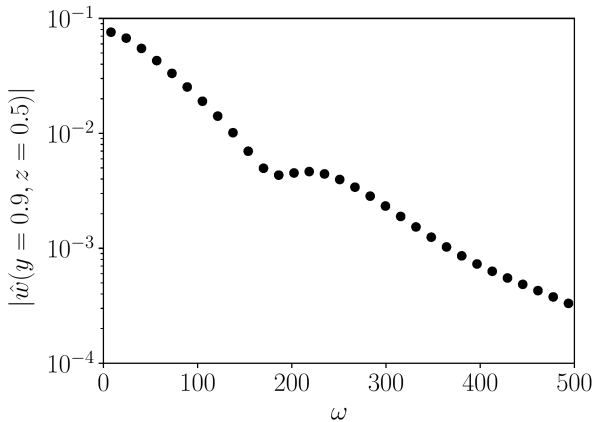}% Here is how to import EPS art
\caption{\label{fig:spectre_De130} 
  Cross stream velocity $w$ spectrum, $De=130$. The dynamics is dominated  by the first mode, which corresponds to a period $T=0.7764$. Coherently with the observations and the symmetry of the problem, the zero mode vanishes in the spectrum, since the average $w$ is vanishing  }
\end{figure}

The spectrum  in figure \ref{fig:spectre_De130} shows  the amplitudes of the Fourier modes of the cross stream velocity $w$ measured at $y=0.9045,z=0.5$  at $De=130$, thus in the periodic regime. The spectrum was obtained by means of the FFT on a time series collected over 5 oscillation periods. The rapid exponential decrease of the amplitude of the modes with increasing pulsations indicates that the  temporal dynamic in the periodic regime is dominated by the first oscillating mode.

\begin{figure*}
\includegraphics[width=0.8\textwidth,keepaspectratio]{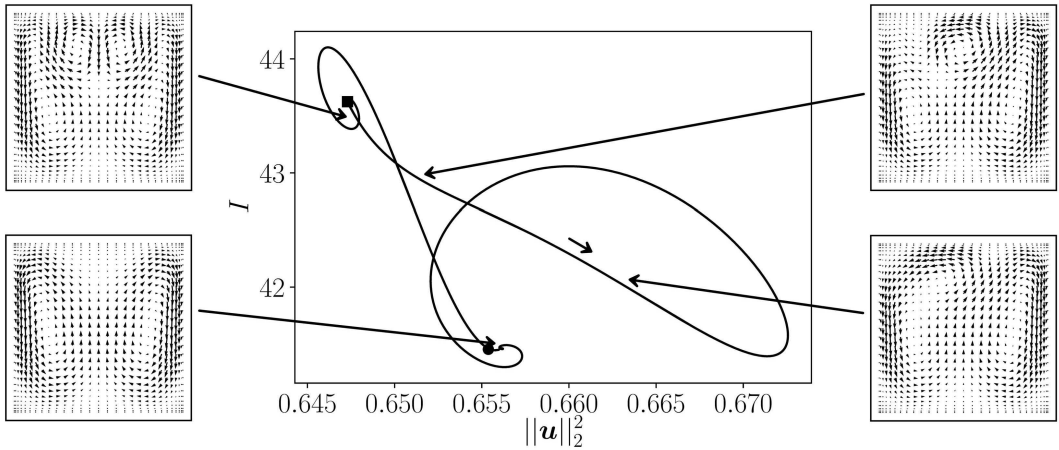}
\caption{\label{fig:phasePortraitDe130} 
 Injection-velocity norm phase diagram at $De=130$. The system follows a periodic orbit passing close to the 2 stationary solutions, indicated by the solid circle ($I \approx 41.5,||\boldsymbol{u}||_2^2 \approx 0.655$) and the solid square (($I\approx43.5,||\boldsymbol{u}||_2^2 \approx 0.647$)). The long arrows indicate the instantaneous states of the system. The flow evolves following the small arrow: from the weak 4-cells, symmetric state to the symmetric strong 4-cells, when symmetry is broken a kinetic energy burst is observed. The system then returns to a weak 4-cells state regaining its symmetry. }
\end{figure*}

The trajectory approaches and leaves periodically the vicinity of two equilibrium states, where the fluid's dynamics slows down and its velocity fields resembles that of the equilibrium. These two equilibria are symmetric with respect to the spanwise direction (z) and are sometimes referred to as the 2 and 4 vortices  solutions\cite{cheng1969laminar,Joseph1975,winters1987bifurcation,YANASE_dualsolutionsPipe_1989}. The vector fields of these two solutions are shown on the left part of figure \ref{fig:phasePortraitDe130}. 
However the 2 or 4 vortices classification can be misleading, because for $De>122.5$  the 2 vortices solution presents in truth 4 cells, even if the additional cells are much less intense than those observed in the 4 cell case. This was also observed by \citet{Watanabe2013} who named the solutions the weak and the strong 4 vortices solutions. We will henceforth use this designation. 

These streamwise vortices are the result of two different physical mechanisms. 
First, for confined curved flow, like in duct or pipe, the  Coriolis acceleration  ($-De\cdot u^2$ which is non-homogeneous in spanwise direction) is in equilibrium with the radial  pressure gradient ($\partial p/\partial y$). As a consequence there is a spanwise pressure gradient ($\partial p/\partial z\ne 0$) which induces spanwise flow ($w \ne 0$). This  mechanism is responsible of the so-called Dean vortices \cite{dean1927,Dean1928, dean_hurst_1959} which correspond to the 2-vortices steady solution mentioned above.
Secondly, for a flow over a concave wall, when the wall-normal pressure gradient ($\partial p /\partial y$) is strong enough, it can give rise to  streamwise  counter-rotating vortices, this phenomenon is known as the  Taylor-G\"ortler instability.
The spanwise wavelength increases with the Reynolds number. The concomitance of these two mechanisms leads to the complexity mentioned above, with the observation of 4 strong or weak cells \cite{Watanabe2013}, and the possibility to observe 6 or more cells for higher Dean number.
%The two additional vortices are produced by the presence of a pressure gradient at the outer wall that balances the centrifugal force given by the $u^2$ term. 
%As seen above the radial  pressure gradient ($\partial p /\partial y$) plays a key role, when it is strong enough, it can give rise to a negative velocity $v$ at the centerline at the outer wall, producing a new, smaller pair of counter rotating vortices. 

Both the weak and strong 4 vortices  steady  solutions are observed even for subcritical Dean numbers. The strong 4-vortices solution is linearly unstable to non-symmetric perturbations (and stable otherwise)\cite{winters1987bifurcation,bara_nandakumar_masliyah_1992}, whereas for $De<De_c$  the weak 4 vortices solution is linearly stable.

This fact can be shown by performing  linear stability analysis of those two equilibrium states at $De_c$. The NS equations are linearized around either of the steady state solutions and  the perturbations can be expressed in the form 
\begin{equation}
\label{eqn:expon_linstab}
{q'}(y,z,t)=\sum_{n}Q_n(y,z)e^{\sigma_n t}
\end{equation}
where $Q_n$ is the $n^{th}$ eigenmode associated to the eigenvalue $\sigma_n$. The real and imaginary parts of $\sigma$ are the growth rate and the pulsation, respectively.
The equilibrium state is unstable if the real part of at least one $\sigma_n$ is positive and stable otherwise.
The asymptotic behaviour is determined by the least stable mode. 
Numerically this least stable mode is calculated with
\begin{equation}
\sigma=  \frac{1}{ \int_{S}^{}  {  q  }^*q~dS} ~  \int_{S}^{}  {q}^*\derp{ q}{t}~dS
\end{equation}
(the  $* $ denotes the complex conjugate) until convergence in the growth rate.
The linear simulations are performed by imposing 
symmetry or anti-symmetry.
 Table \ref{tab:stability_stronweak128.32} shows the  eigenvalues  of the most amplified mode for the weak and strong 4-vortices solutions and for symmetric and anti-symmetric perturbations. Alike in the literature previously mentioned, the strong 4 vortices solution is stable to symmetric perturbations and unstable to anti-symmetric ones. The weak 4-vortices solution is linearly stable to any perturbation in the subcritical regime.

Thus, when the system is non-symmetrically perturbed for $De<De_c$, it evolves  close to the strong 4-vortices solution, which is a saddle point, thus the trajectory is ejected and the system eventually returns to the weak 4 cell solution, which is the global attractor.
This is the reason why the latter is the observed solution in direct numerical simulation or experiments in the subcritical regime. 
%This also means that the instability is not given by  the presence of the additional vortices, as a 4 vortices solution (the weak one) can be linearly stable. 
When the system is non-symmetrically perturbed for $De>De_c$, the weak 4-vortices solution also becomes linearly unstable, a perturbation will therefore initiate a cycle  between the two solutions. The periodic orbit of the system at $De=130$ is depicted in figure \ref{fig:phasePortraitDe130} in a state space diagram spanned with the energy injection and squared velocity 2-norm.   
The system follows the direction indicated by the small arrow in the center of the figure. Starting in the proximity of the weak 4 vortices solution (indicated with the black circle at the bottom left) the flow evolves towards the strong 4 vortices solution (indicated with the black square at the top left). The system  then loses its symmetry, which is consistent with the instability to asymmetric perturbations of this solution. The two upper vortices collapse into one lateral vortex, producing a   burst. The system then  rapidly returns to a symmetric state, namely  the weak 4 vortices solution.

The trajectory is attracted by the stable manifold (symmetric) of the strong 4-vortices equilibrium,
it then leaves this region along the unstable manifold (antisymmetric) and heads to the second equilibrium state (weak 4-vortices).
If $De>De_c$, then the trajectory follows the second heteroclinic orbit along the unstable manifold (symmetric) of the weak 4-vortices equilibrium,
it then heads to the strong 4-vortices equilibrium state, closing the cycle.
These two heteroclinic orbits between the two equilibria constitute a heteroclinic cycle which can be suppressed by imposing spanwise symmetry.

% The weak 4 vortices solution is also present at $De<De_c$, in fact both the strong and the weak 4 vortices solutions are present for slightly subcritical Dean numbers.
% The strong 4 vortices solution is linearly unstable to asymmetric perturbations, and stable to symmetric perturbations\cite{winters1987bifurcation,Wang2005}, whereas the weak solution is linearly stable for any perturbation, for $De<De_c$. 

\begin{table}
\caption{\label{tab:stability_stronweak128.32} 
Eigenvalue of the most unstable eigenmode of the strong and weak 4-vortices solutions at $De_c=128.32$, obtained by means of the linear stability analysis.  The convergence was stopped when the change in value dropped below $10^{-4}$. 
The eigenvalues can be considered as real, in other words the pulsation is numerically zero.
%The strong 4-vortices solution is stable to symmetric perturbations and stable to anti-symmetric perturbations, the weak solution is unstable to antisymmetric ert 
} 
\begin{ruledtabular}
\begin{tabular}{lccccc}
   & Symmetry & Antisymmetry  \\ \hline
 Strong 4 vortices   &   $-40.20+0i$        & $+44.215-4.5\cdot10^{-4}~i$        \\   
 Weak 4 vortices    &   $-0.540-4.6\cdot10^{-4}~i$         & $-22.371-4.6\cdot10^{-4}~i$         \\   
\end{tabular}
\end{ruledtabular}
\end{table}

\subsection{High Dean regime, $\mathbf{De= 10^4}$}
\begin{figure}
\includegraphics[width=0.45\textwidth,keepaspectratio]{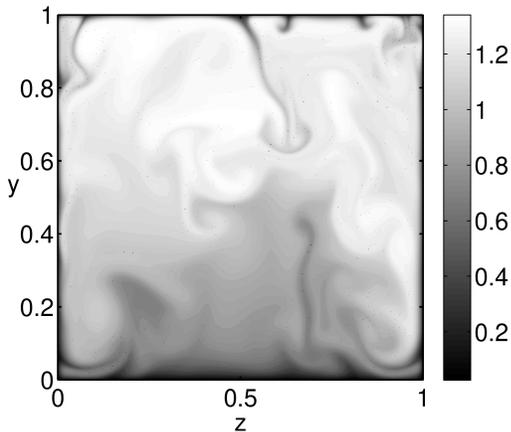}
\caption{\label{fig:De1e4_isoU}
Isocontour of the streamwise velocity $u$. Snapshot of simulation at $De= 10^4$
}
\end{figure}

The application of dynamical systems ideas to fluid mechanical problems has provided much insight into the ways fluids transit from a laminar state to a turbulent one. 
The identification of the different routes to chaos suggests they can also be routes to
turbulence that lead from a laminar flow state through various bifurcations to a turbulent attractor.
In this section we introduce a relatively high Dean number simulation, at $De= 10^4$. The numerical parameters needed to be adapted $Ny=Nz=501$, $\Delta t=10^{-8}$. Figure \ref{fig:De1e4_isoU} shows a snapshot of  isocontours of the streamwise velocity $u$. As can be seen the topology changes drastically as compared to the 2 or 4 cells solutions observed so far at much lower  Dean number (figure \ref{fig:phasePortraitDe130}).

The solid no-slip walls are source of filament structures\cite{2009_AMR_Clercx}, which are advected away from the walls and  disrupt  the Dean vortices. While Dean vortices fill the whole domain, in other words they scale with the duct height $h$, the filament structures scale with the vorticity thickness at the wall.  The ratio between these two length scales is related to the Dean number, thus for high Dean number we observe a  multiscale dynamic leading to a complex spatio-temporal pattern. The 2D restriction is too strong to qualify this flow as realistically turbulent, nonetheless the richness of this minimal model permits to investigate turbulence within a dynamical system framework (see the recent discussion on the subject in Ref \onlinecite{hof2017searching} and references therein). 

% % COMMENTED BECAUSE LATEX DID NOT COMPILE

% \begin{figure}
% \includegraphics[width=0.45\textwidth,keepaspectratio]{De1e4_Utau.eps}
% \caption{\label{fig:De1e4_Utau} 
% Space-time diagram of the shear velocity at $y=0.5???????$.  $De=1\cdot 10^4$
% }
% \end{figure}

\section{Conclusion and perspectives}
In this paper we developed a simplified model for the flow in a weakly curved square duct, in line with the work by \citet{dean1927}. However,  while the original Dean model is based on a truncation at $\mathcal{O}(h/R)$,  the present model is truncated at $\mathcal{O}(\sqrt{h/R})$, the square root  comes from the use of two characteristic velocities. 
The simplified model is Cartesian and is governed by one control parameter, the numerical solution is therefore much simpler  compared to the full Navier Stokes equations in polar coordinates.
We compared the results with the literature \cite{winters1987bifurcation,Wang2004},  finding a very good agreement in the bifurcation points and the oscillation period.

Moreover the model allowed us to investigate the periodic regime of the system in terms of both bifurcations and physical mechanisms, in addition, it was shown that the observed periodicity results of a  heteroclinic cycle constituted by two heteroclinic orbits between the two equilibria: the weak and the strong 4-vortices solutions.
The trajectory is backward asymptotic to one equilibrium state and forward asymptotic to the other and reciprocally on the way back closing a heteroclinic cycle.

The model is an example of deterministic chaos in fluid flows, as already seen in Rayleigh-B\'enard convection or  in  Taylor-Couette flow between two rotating cylinders. Nonetheless these flows differ by the coupling term generating the streamwise vortices. In the present work this term is quadratic ($-De ~ u^2$) while it is linear in rotating or convective cases. This difference deserves investigations, and that was the original objective of the present paper. In that purpose,  the complete program to solve 2D Navier-Stokes equations is made freely available.

Although presented in 2D in this paper, the present model was originally developed in 3D. In a future work we plan to extend our analysis to more complex flows, namely using the 3D version of the present model. A step was made in that direction by \citet{sankar1988} for steady flows, and it appears interesting to continue towards more complex dynamics induced by  turbulent inflow boundary conditions. In order to investigate the paradoxical effect of curvature: stabilising turbulence \cite{Sreenivasan1983,taylor1929,white_1929,vester_2016Review} on one hand and generating peculiar streamwise vortices on the other hand.

% - prolongate Dean model compared to full model wang, winters

% - results in good agreement: crit dean, period
% - cartesian grid, 1 control parameter
% - extension to more complex flows: 3d, turbulent inflow
%  \citet{sankar1988} obtained a value of $128.125$ using a 3D stationary, parabolic  formulation. 
% - While the original Dean model is based on a truncation at order h/R the present model is truncated at (h/R)**0.5, in conjunction with the 2-scale velocity assumption. 
% - the model is an example of deterministic chaos in fluid flows, as already seen in Rayleigh Benard convection or in rotating systems, nonetheless the main difference is in the coupling term responsible of the streamwise vortices, in the present work this term is quadratic (De u**2) while it is linear in rotating or convective cases, this slight difference deserves investigation, this was the original purpose of the paper and the script is proposed for whom is interested in going deeper

%\begin{acknowledgments}
%We wish to acknowledge the support of the author community in using
%REV\TeX{}, offering suggestions and encouragement, testing new versions,
%\end{acknowledgments}
\begin{table}
\caption{\label{tab:calculation_convergence_De150} Comparison between two calculations at $De=150$, the quantities reported are averaged over one time  period. The coarser space ($N$, number of grid points) and time discretisations (third line of the table) are those used in the rest of the paper, the finer simulation is used as a benchmark. Very little differences are observed in the results: the period $T$ varies of a quantity compatible with the $dt$ itself, the average pressure gradient $dP/dx$ and the streamwise and span wise dissipations $\varepsilon_u$,$\varepsilon_{vw}$ exhibit very limited differences. The differences between the  streamwise pressure gradient (equal to the energy injection if $U_b=1$) and the streamwise dissipation indicate the level of convergence of the respective calculations.   }
\begin{ruledtabular}
\begin{tabular}{lccccc}
 $N_y\times N_z$ & $\Delta t$              & $T$       & $dP/dx$    & $\varepsilon_u$ & $\varepsilon_{vw}$ \\ \hline
 $101 \times 101$     & $1\cdot10^{-6}$   & $0.20132$ & $-44.5091$ & $44.5066$       & $11.43055$         \\
 $31 \times 31 $      & $1.2\cdot10^{-5}$ & $0.2013$  & $-44.513$  & $44.5130$       & $11.43557$      
\end{tabular}
\end{ruledtabular}
\end{table}

\appendix
\section{The code}
\label{sec:appendix}
Equations (\ref{eqn:NS_NONdimensional_minimalmodel_xyz}) are solved using a  solver based on  Chebyshev collocation method. 
The  time  marching  combines
a  fourth-order  Adams-Bashforth  scheme  and  a  fourth-order  backward  differentiation
scheme, with the viscous term treated implicitly \cite{Ascher1995}.

The   divergence-free   condition   is   achieved   with   the   prediction-projection  scheme by Chorin and Temam (for an exhaustive review about projection methods see Ref \onlinecite{GUERMOND20066011}).
Pressure is computed using polynomials of 2 orders less than those used for the velocities,  in order for the pressure field to remain unpolluted of spurious modes. The same grid is used for pressure and velocity fields and  no pressure boundary condition is required.  For further details on the accuracy and stability of the method see Ref \onlinecite{BOTELLA1997107}. 
This numerical method was found to be accurate enough to compute the subcritical transition to turbulence in square duct flows\cite{BiauPhilosop2009}.
Most of  the results have been obtained with a grid composed of $Ny\times Nz=31\times 31$ Chebyshev points, with a time step ($\Delta t$) of $1.2\cdot 10^{-5}$. A quantitative comparison with a simulation performed using finer time step and grid can be found in table \ref{tab:calculation_convergence_De150}. The error in the oscillation period is of the same order of the time-step, the differences in the streamwise pressure gradient and streamwise and cross-stream dissipations averaged on one time period are below $5\%$, despite a more than ten-fold increase in the number of grid points and a 12-times finer time-step. The agreement  between the average  dissipation and pressure gradient in the streamwise direction testifies  the convergence of the calculations. 

Pages \pageref{page:code_main} and \pageref{page:code_functions} contain the source code used to generate the data shown in the present paper. This version of the code mostly contains the calculating core of the program, while the most part of the post-processing  was not included for clarity. Some quantities of interest are however  added for validation purposes (the streamwise pressure gradient and the dissipations in streamwise direction and the dissipation in cross-flow directions $x$ and  $y,z$-). The reader is encouraged to run the program and compare the results obtained with those reported in table \ref{tab:calculation_convergence_De150}. The simulation is initialized with the laminar solution plus a random velocity field, thus the initial transient shows a little sensitivity from run to run.
The 'functions' file is composed by four methods, inserted in a class for practicality.
The 'constBulk' method uses a bisection algorithm to iteratively calculate the streamwise pressure gradient required to maintain the target streamwise bulk velocity ($U_b=1$ in the present paper).
The 'Chebyshev' method is used to define the grid, the derivation matrices for velocity and pressure and the integration weights.  The other two methods are auxiliaries for the 'Chebyshev' method.
The first part of the 'main' file serves as  initialization, here we define the operators for derivations and integrations in $y$ and $z$, the extrapolation ($c_i,d_i$) and the time integration ($a_1,b_i$) coefficients. Finally we set the initial condition with the laminar flow solution in a straight duct ($Ulam$), with some additional random noise.

The following part is the actual time loop. In the first part we compute the right-hand-side term (explicit) of the equations, then the solution is updated for the next time-steps before calculating the new predicted velocity field. This is then corrected to enforce the divergence free condition following the prediction-projection method by Chorin and Temam.
%\citet{CHORIN196712}. 
%This correction is not applied to the streamwise velocity since $\partial p / \partial x=0$. 
As a last step of the loop the  streamwise pressure gradient is calculated using the bisection method mentioned above, and the streamwise velocity is corrected as well.

The code is available for download in the online version of the paper, but can also be copy-pasted from the present $.pdf$ file and saved in two files: $main.py$ and $functions.py$, in the same folder. In this case particular attention must be paid on accurately copying the  indentations, which are essential in a python program. The present results were obtained by running the program on a linux machine using the Anaconda distribution of python version 3.6.4. The code also works on python 2.7 (a test was run on the Anaconda distribution of python 2.7.16).

\onecolumngrid
 \clearpage

% \newpage
 \subsection{Main}

 \label{page:code_main}
\scriptsize{
\begin{Verbatim}
import numpy as np
from functions import functions_cl as ff

De=150  ;   dt=1e-5

Ny = 31; Nz = Ny
#  Chebyshev collocation method, operators definition
cc=ff()
[Y,D1,D2,Dpy,WY]=cc.Chebyshev(Ny)
Y=(1-Y)/2; WY=-WY/2; D1=D1*2
y = Y[1:-1]; Wy = WY[1:-1]; Dy = -D1[1:Ny-1,1:Ny-1]; D2y = 4*D2[1:Ny-1,1:Ny-1]; Dpy=-2*Dpy; Ay = np.dot(Dy,Dpy)
[Ey_u,Qy_u] = np.linalg.eig(D2y); invQy_u=np.linalg.inv(Qy_u);  Ey_u=(np.tile(Ey_u, [Nz-2,1])).T
[Ey_p,Qy_p] = np.linalg.eig(Ay);  invQy_p=np.linalg.inv(Qy_p);  Ey_p=(np.tile(Ey_p, [Nz-2,1])).T
Dz=np.transpose(Dy); D2z=np.transpose(D2y); Dpz=np.transpose(Dpy); Az=np.transpose(Ay);
Qz_u=np.transpose(Qy_u);  invQz_u=np.transpose(invQy_u);  Ez_u=np.transpose(Ey_u);
Qz_p=np.transpose(Qy_p);  invQz_p=np.transpose(invQy_p);  Ez_p=np.transpose(Ey_p);

d1 = 4.;d2 = -6.;d3 = 4.;d4 = -1.                              # extrapolation coefficients
c1 = 2.;c2 = -1.;c3 = 0.;c4 = 0.                               # lower order for pressure
a1 = 25/12.;b1 = -4.;b2 = 3.;b3 = -4/3.;b4 = 1/4.;sig = a1/dt  # Backward Euler 

# initializing flow field
Ulam =np.dot(np.dot( Qy_u,( np.dot (  np.dot( invQy_u,(-np.ones((Ny-2,Nz-2))) ) , invQz_u)/(Ey_u+Ez_u)) ),Qz_u) 
Ulam = Ulam/( np.transpose(Wy).dot(Ulam).dot(Wy) )
Px=Wy.dot(Dy[0,:].dot(Ulam))*4
uf1=np.copy(Ulam)+np.random.rand(Ny-2,Ny-2)*1e-2; uf2=np.copy(uf1);uf3=np.copy(uf1);uf4=np.copy(uf1);
vf1=np.zeros((Ny-2,Ny-2)); vf2=np.copy(vf1); vf3=np.copy(vf1); vf4=np.copy(vf1)
wf1=np.zeros((Ny-2,Ny-2)); wf2=np.copy(wf1); wf3=np.copy(wf1); wf4=np.copy(wf1)
pf1=np.zeros((Ny-2,Ny-2)); pf2=np.copy(pf1); pf3=np.copy(pf1); pf4=np.copy(pf1)

timeEnd=1   ;k=0          ;U=np.zeros((Ny,Nz))
time=[0]    ;Px_time=[0]  ;V=np.zeros((Ny,Nz))
dissip_u=[0];dissip_vw=[0];W=np.zeros((Ny,Nz))
# ------------------------
#   starting time loop
while time[-1] < timeEnd:
    k+=1
    u=d1*uf1+d2*uf2+d3*uf3+d4*uf4
    v=d1*vf1+d2*vf2+d3*vf3+d4*vf4
    w=d1*wf1+d2*wf2+d3*wf3+d4*wf4
    p=c1*pf1+c2*pf2+c3*pf3+c4*pf4
    # -----  calculating right hand side  ----- #
    rhs_u = -( b1*uf1+b2*uf2+b3*uf3+b4*uf4 )/dt +De*(-v*(Dy.dot(u)) -w*(u.dot(Dz)))
    rhs_v = -( b1*vf1+b2*vf2+b3*vf3+b4*vf4 )/dt +De*(-v*(Dy.dot(v)) -w*(v.dot(Dz)) +u**2) - Dpy.dot(p)
    rhs_w = -( b1*wf1+b2*wf2+b3*wf3+b4*wf4 )/dt +De*(-v*(Dy.dot(w)) -w*(w.dot(Dz)))       - p.dot(Dpz)
    # ---------------  update  ---------------- #
    uf4=np.copy(uf3); uf3=np.copy(uf2); uf2=np.copy(uf1);
    vf4=np.copy(vf3); vf3=np.copy(vf2); vf2=np.copy(vf1);
    wf4=np.copy(wf3); wf3=np.copy(wf2); wf2=np.copy(wf1);
    pf4=np.copy(pf3); pf3=np.copy(pf2); pf2=np.copy(pf1);
    # -------------  prediction  ------------- #
    uf1 = Qy_u.dot( (invQy_u.dot(rhs_u).dot(invQz_u)) / (sig-Ey_u-Ez_u) ).dot(Qz_u)
    vf1 = Qy_u.dot( (invQy_u.dot(rhs_v).dot(invQz_u)) / (sig-Ey_u-Ez_u) ).dot(Qz_u)
    wf1 = Qy_u.dot( (invQy_u.dot(rhs_w).dot(invQz_u)) / (sig-Ey_u-Ez_u) ).dot(Qz_u)
    # -------------  projection  ------------- #
    div = Dy.dot(vf1)+wf1.dot(Dz)
    RHSP = sig*div + Ay.dot(p) + p.dot(Az)
    pf1 = Qy_p.dot( ((invQy_p.dot(RHSP)).dot(invQz_p))/(Ey_p+Ez_p)).dot(Qz_p)
    vf1 -=  1.0/sig*Dpy.dot((pf1-p))
    wf1 -=  1.0/sig*(pf1-p).dot(Dpz)
    Px = ff.constBulk(Px,Qy_u,invQy_u,Ey_u,Qz_u,invQz_u,Ez_u,sig,Wy,rhs_u)
    uf1 = np.dot( Qy_u,( ( np.dot( invQy_u,(rhs_u-Px ) ).dot(invQz_u) )/(sig-Ey_u-Ez_u) ) ).dot(Qz_u)
    
    if( k % 200  ) == 0:
      time.append(k*dt) ; Px_time.append(Px)
      # ------------------------  calculating dissipation  ------------------------ #
      # ----  derivative is non zero on boundary -> adding the implicit point  ---- #
      U[1:-1,1:-1]=u;   V[1:-1,1:-1]=v;   W[1:-1,1:-1]=w
      dissip_u.append(WY.dot((D1.dot(U))**2 + (U.dot(D1.T))**2).dot(WY) )
      dissip_vw.append(WY.dot((D1.dot(V))**2+(D1.dot(W))**2+(V.dot(D1.T))**2+(W.dot(D1.T))**2).dot(WY))
      # --------------------------------------------------------------------------- #
      divafter=Dy.dot(v)+w.dot(Dz)
      print('time: '+str('%0.3f'%time[-1])+ ' str-wise dissip:'+str('%0.3f'%dissip_u[-1])+
              ' div before/after proj: ' +str('%0.3e'%np.abs(div).max()) +'/'+
              str('%0.3e'%np.abs(divafter).max()) +' u bulk error:'+str('%0.3e'%(Wy.dot(uf1).dot(Wy)-1)))

print('average stream-wise dissipation: '+str('%0.5f'%(np.asarray( dissip_u [100:] ).mean())))
print('average stream-wise press grad : '+str('%0.5f'%(np.asarray( Px_time  [100:] ).mean())))
print('average span-wise   dissipation: '+str('%0.5f'%(np.asarray( dissip_vw[100:] ).mean())))
\end{Verbatim}}
\clearpage
% %\VerbatimInput{figures/simplified_code.py}
\clearpage
\subsection{Functions}
\label{page:code_functions}
\scriptsize{
\begin{Verbatim}
import numpy as np
class functions_cl:
    # ------------------------------------------
    # generation of grid and derivation matrices
    @staticmethod
    def cheb(N):
        if N == 0:
            D = 0
            x = 1
            ll=(D,x)
            return ll
        x = np.cos( np.pi*np.arange(N+1)/float(N) )
        c=np.ones(N+1)
        c[[0,-1]]=2
        c=c*(-1)**(np.arange(N+1))
        X = np.transpose(np.tile(x,[N+1,1]) )
        dX = X-np.transpose(X)
        # off-diagonal entries      
        D  =  np.outer(c ,1/c) / (dX+np.eye(N+1))
        # diagonal entries 
        D  = D -  np.diag(np.sum( np.transpose(D),0 ))
        return D,x
    @staticmethod
    def clencurt(N):
        # CLENCURT  nodes x (Chebyshev points) and weights w
        #           for Clenshaw-Curtis quadrature
        theta = np.pi*np.arange(N+1)/N
        w = np.zeros(N+1)
        ii = range(1,N)
        v = np.ones(N-1)
        if np.mod(N,2)==0:
            w[0] = 1./(N**2-1)
            w[N] = w[0]
            for k in np.arange (1,N/2):
              v = v - 2*np.cos(2*k*theta[ii])/(4*k**2-1)
            v = v - np.cos(N*theta[ii])/(N**2-1)
        else:
            w[0] = 1./N**2
            w[N] = w[0]
            for k in np.arange (1,(N-1)/2+1):
                v = v -    2*np.cos(2*k*theta[ii])  /  (4*k**2-1)
        w[ii] = 2.*v/N
        return w

    def Chebyshev(self,Ny):
        [Dy,y] = self.cheb(Ny-1)
        D2y = np.dot(Dy,Dy)
        Wy = self.clencurt(Ny-1)
        #Wy = ChebyshevCl.clencurt(Ny-1)
        #  see Botella, Dpy : derivative matrix without implicit BC
        c = (np.sin(  np.pi*(np.arange(1,Ny-1))/(Ny-1)    )**2) *(-1) ** np.arange(1,Ny-1)
        X = np.transpose(np.tile(y[1:Ny-1],[Ny-2,1]))
        dX = X-np.transpose(X)
        # off-diagonal entries                 
        D0 = np.outer((1./c),c)/(dX+(np.eye(Ny-2)))
        # diagonal entries
        Dpy = D0 - np.diag(np.sum(D0,1))
        return (y,Dy,D2y,Dpy,Wy)
    
    # -----------------------------------
    #   Bisection method for mean flow
    @staticmethod
    def constBulk(Px,Qy_u,invQy_u,Ey_u,Qz_u,invQz_u,Ez_u,sig,Wy,rhs_u):
        Px1=Px-.2;  uf1 = np.dot( Qy_u.dot( ( invQy_u.dot(rhs_u-Px) ) ).dot(invQz_u)/(sig-Ey_u-Ez_u) , Qz_u); 
        Q1=np.dot(np.dot(Wy,uf1),Wy);
        Px2=Px+.2;  uf1 = np.dot( Qy_u.dot( ( invQy_u.dot(rhs_u-Px) ) ).dot(invQz_u)/(sig-Ey_u-Ez_u) , Qz_u); 
        Q2=np.dot(np.dot(Wy,uf1),Wy);
        iter=1; Q=(Q2+Q1)/2;
        while abs(Q-1) > 1e-14:
            Px = (Px1+Px2)/2;
            uf1 = np.dot( Qy_u.dot( (invQy_u.dot((rhs_u-Px)) ).dot(invQz_u) / (sig-Ey_u-Ez_u) ) , Qz_u )
            Q=np.dot( np.dot(Wy,uf1),Wy )
            if Q>1:
                Px1=Px
            else:
                Px2=Px
            iter+=1
            if iter >=300:
                break
        return Px
\end{Verbatim}}
\clearpage
 \twocolumngrid

%\nocite{*}
\bibliographystyle{h-physrev}
\bibliography{aipsamp}% Produces the bibliography via BibTeX.

\end{document}